\documentclass{aa}%
\usepackage{graphicx,natbib,ulem}
\usepackage{color}
\usepackage{amsmath}
\usepackage{txfonts}

\begin{document}
  \titlerunning{W 50 and SS 433}
\authorrunning{Bowler \& Keppens}
   \title{W 50 and SS 433}

   \subtitle{}

   \author{
          Michael G.\ Bowler \inst{1} \and Rony Keppens\inst{2}
          }

  \offprints{M.G. Bowler \\  \email{michael.bowler@physics.ox.ac.uk}}
   \institute{University of Oxford, Department of Physics, Keble Road,
              Oxford, OX1 3RH, UK \and Centre for mathematical Plasma Astrophysics, Department of Mathematics, KU Leuven
Celestijnenlaan 200B, 3001 Heverlee, Belgium}
   \date{Received; accepted}

% \abstract{}{}{}{}{} 
% 5 {} token are mandatory
 
  \abstract
   {The Galactic microquasar SS\,433 launches oppositely-directed jets at speeds approximately a quarter of the speed of light. These appear to have punched through and beyond the supposed supernova remnant shell W 50. The problems with this interpretation are: (i) The precessing jets have somehow been collimated before reaching the shell; (ii) Without deceleration, only recently launched jets would have reached no further; and (iii) Certain features in the lobes are moving slowly or are stationary.}
  {Hydrodynamic computations have demonstrated that for at least one set of parameters describing the ambient medium, jets that diverge and precess are both decelerated and collimated; the conformation of W 50 could then have been sculpted by the jets of SS 433. However, the parameters adopted for density and pressure in these computations are not consistent with observations of jets at a few years old; nor do they represent conditions within a supernova remnant. Our aim is to investigate whether the computations already performed can be scaled to a realistic W 50.}
   {We find simple and physically based scaling relations. The distance to collimation varies inversely with the square root of the pressure of the ambient medium and the speed with which the head of a collimated jet propagates scales with the square root of the temperature. We extrapolate the results of the hydrodynamic computations to lower densities and pressures.} 
  {The jets of SS 433, launched into an ambient medium of pressure $\sim 10^{-9}$ erg cm$^{-3}$ and temperature $\sim 10^{8}$ K, within a supernova remnant, could be responsible for the characteristics of W 50. The precessing jets are collimated within $\sim$ 10 pc and the head of the resulting cylindrical jet propagates slowly.}
  % conclusions heading  (optional), leave it empty if necessary 
{The problems of relating W 50 to SS 433 may now be solved.}


   \maketitle
%
%________________________________________________________________

\section{Introduction}
\subsection{RA 19$^{h}$ 12$^{m}$ Dec 4$^{\circ}$ 59$^{'}$}
The above  coordinates (J2000) give the location on the sky of two very different objects, each remarkable in its own right and each remarkable in its evident relationship with the other. One is the microquasar SS 433 that emits opposite jets of baryonic material at speeds $\sim$ 0.26c, more or less continuously. The other is the radio nebula W 50, remarkable for its size, brightness and peculiar morphology. To east and west of an approximately spherical shell that might be a supernova remnant are extended lobes on an axis shared with the jets of SS 433. SS 433 sits at the core of W 50, the nebula and the microquasar being at the same distance, 5.5 kpc, from our telescopes. It was suggested a long time ago that the jets have extruded through the expanding shell of the supernova
 remnant that gave birth to the compact object. 
\subsection{SS 433}
\begin{figure}[ht]
      \centering
      \includegraphics[width=8cm]{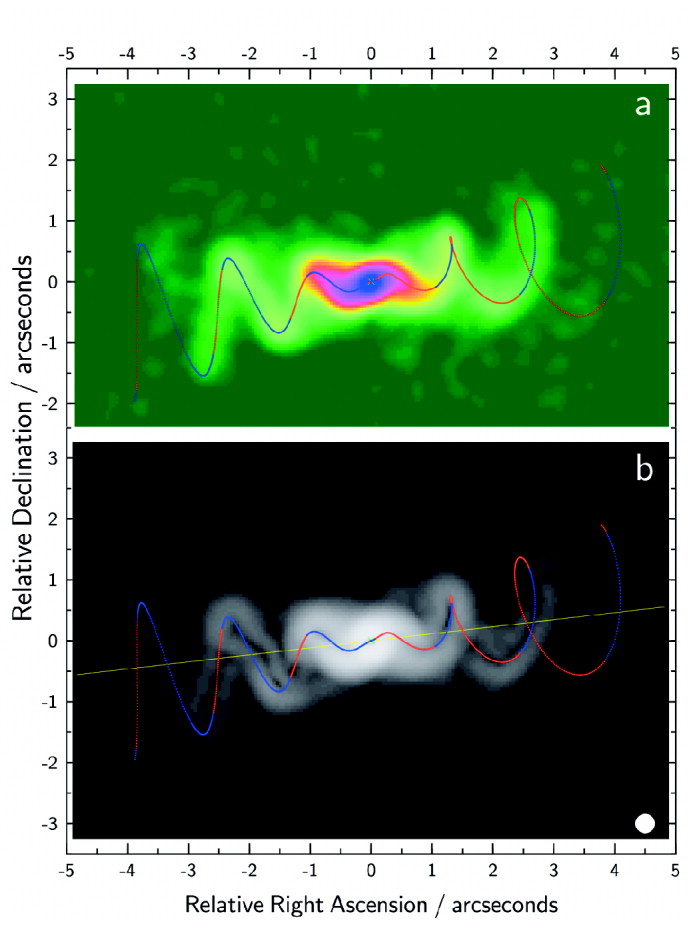}
      \caption{Two renderings of VLA data showing several precessional periods of the jets of SS 433. In the lower panel the ridge line is emphasised, superimposed are the traces of the simplest kinematic model. From~\citet{BlundellBowler2004}.}
\label{fblundell}
    \end{figure}

Object number 433 in the catalogue of stars with strong emission lines compiled by Stephenson \& Sanduleak was the first microquasar to be identified. The emission lines were found to have very large Doppler shifts, varying periodically with time. Analysis rapidly showed that the emission lines matched remarkably well a model in which they originated in two oppositely directed jets of gas, particularly strong in H$\alpha$, emitted at 0.26$c$. These jets precess and make an angle of 20$^{\circ}$ to the precession axis, drawing a helix of expanding radius across the sky, Fig.~\ref{fblundell}~\citep{BlundellBowler2004}. The period of precession is 162 days and the axis nods with a nutation period of approximately 6 days, the system being binary with orbital period 13.08 days. The compact object is super-Eddington accreting from a comparatively normal star. The jets fade rapidly in the visible spectrum, but have been mapped in radio out to 4$^{''}$, corresponding to 0.1 pc at distance of 5.5 kpc. An example of deep imaging is shown in Fig.~\ref{fblundell}, drawn from~\citet{BlundellBowler2004}. The image is augmented with the trace of the precessing jets and the projection on the sky of the precession axis. The cone angle (twice the angle to the precession axis of a jet) is easily seen to be $\sim$40$^{\circ}$. The east and west jets look different; from this relativistic aberration their speed and hence the distance to SS 433 was extracted independent of optical Doppler shift~\citep{BlundellBowler2004}.

 Even as a microquasar SS 433 is unique. It is the only known Galactic microquasar where line spectra from atomic transitions show baryonic matter to dominate in the jets, the jets being close to continuous rather than intermittent and accretion highly super-Eddington.  Extensive reviews of the SS 433 system may be found in~\citet{Margon1984} and \citet{Fabrika2004}, recently it has emerged that it may be the only Ultra Luminous X-ray source in the Galaxy~\citep{Fabrikaetal2015}.

\subsection{W 50}

\begin{figure}[h]
      \centering
      \includegraphics[width=8cm]{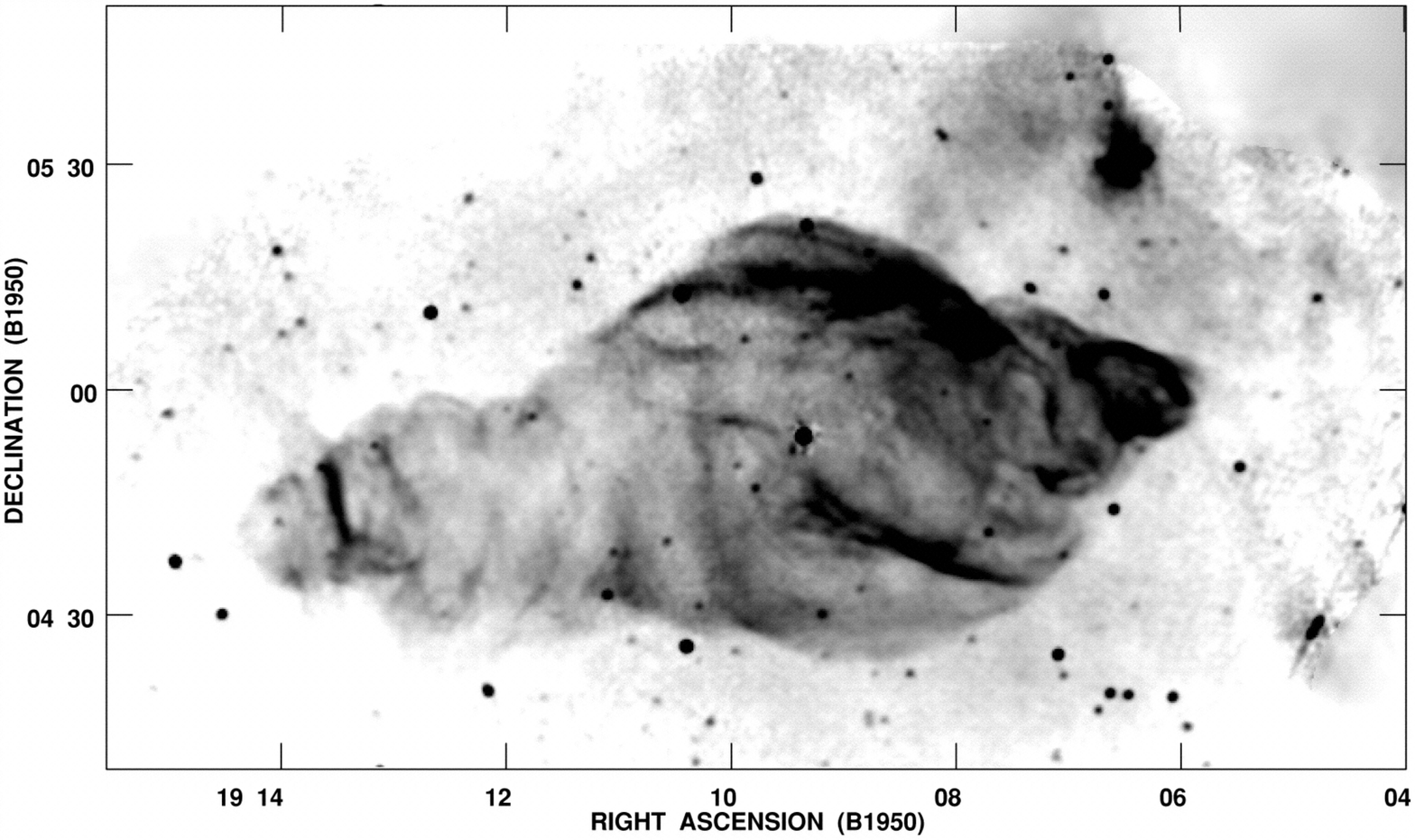}
      \caption{The famous image of W 50 constructed from VLA data and appearing in~\citet{Dubneretal1998}.}
\label{fdubner}
    \end{figure}

    The nebula W 50 spans over 2$^{\circ}$ on the sky and is unusually bright. It has a curious morphology suggestive of a manatee or a seashell - the best- known image is from 1.4 GHz observations with the VLA~\citep{Dubneretal1998}, reproduced here as Fig.~\ref{fdubner}. LOFAR images at approximately 150 MHz have recently been published~\citep{Brodericketal2018} and in Fig.~1 of that paper W50 is displayed on a smaller scale, revealing its Galactic context. Fig.~2 in the same paper is more or less the equivalent of the~\citet{Dubneretal1998} image. This nebula seems to consist of an approximately spherical part that might well be a supernova remnant, but with two protrusions or lobes to east and west. These define an axis aligned with the precession axis of the jets of SS 433, which sits at the core of the spherical component of W50. It is useful to compare Figs.~\ref{fblundell} and~\ref{fdubner} in this context and it should be noted that the scales are a factor approximately 1000 different. The jet structures in Fig.~\ref{fblundell} would fit approximately 16 times into the central spot of Fig.~\ref{fdubner}. The morphology invites an interpretation in terms of a spherical shell through which the jets of SS 433 have punched to extend the protrusions. A major problem with this picture has always been that beyond the supposed SNR shell the divergence from the precession axis of the jets has clearly become less than 10$^{\circ}$, this is clear in both radio and X-ray images. Indeed, a few simple measurements on Fig.~\ref{fdubner} are sufficient to demonstrate this. At the western end of the east lobe little more than a 10$^{\circ}$ precession angle could be tolerated; if the ring at the eastern end is a terminal shock, then even this moderate precession angle is too great. Thus if the jets have always been emitted at 20$^{\circ}$, collimation must have taken place before the expanding helix of Fig.~\ref{fblundell} reached the SNR shell. This is well illustrated in~\citet{Goodalletal2011b}, their Fig.~1. In Fig.~11 of the same paper, the Dubner image is superposed on ROSAT X-ray data: the lobes also appear in X-ray images, a further indication of the influence of the jets in their formation. While the eastern ear extends further than the western, both protrude well beyond the apparent spherical shell presumed to be from the supernova. The extreme tips of the ears from these VLA images are at radii of 121 and 86 pc for east and west ears respectively, while the radius of the spherical shell is about 45 pc at a distance of 5.5 kpc. SS 433 is at a distance of 5.5 kpc~\citep{HjellmingJohnston1981,BlundellBowler2004}; independently~\citet{Lockmanetal2007} found a distance for W 50 bounded by 5.5 and 6.5 kpc. The distance and dimensions of this interstellar manatee are well known; the age is not measured and can only be inferred from models of SNR, assuming that W 50 is indeed a SNR and not some kind of wind-blown bubble.

A recent paper presents many radio images of the full W 50 field of view and also reveals numerous faint H$\alpha$ filaments associated with the central region of W 50, suggestive of SNR shock emission~\citep{Farnesetal2017}.

\section{Collimation?}
    The lobes appear on a scale of approximately 50 - 100 pc, supposedly making visible the jets from SS 433 that disappeared from imaging a thousand times closer in. If we are indeed seeing the same jets in both Fig.~\ref{fblundell} and Fig.~\ref{fdubner}, it is clear that collimation of the precessing jets into something closer to cylindrical form is required. It is necessary to understand how this might happen, even more important to show how it has happened. There have been various attempts, of increasing sophistication as time has gone by and computing power has increased.
To aid the reader on terminology used primarily in numerical modeling, a glossary of geometric terms used to characterise jets is appended.

\subsection{Hollow cones and ambient pressure}

A rather simple analytic model was introduced by~\citet{Eichler1983}. He argued that because of the shortness of the precessional period, precessing jets (of cylindrical cross section) could be modelled as a conical surface propagating outward. Such a hollow cone would cut through the ambient medium and in the simplest picture the pressure of that medium would act equally on the inside and the outside of the hollow cone. Eichler assumed further that the expanding cone would sweep up and clear out the gas from the interior of the cone, making it hollow indeed. The pressure would then act to continually squeeze the rate of radial expansion of the cone until radial expansion is lost, when collimation of a kind would indeed have been achieved. Beyond that, a collapsing cone would lead to a focus. \citet{Eichler1983} showed that this is nearly independent of the original opening angle, and provided an analytic expression for the curved path followed by a jet parcel where inward pressure is balanced by centrifugal forces. It will be relevant later that for given jet characteristics the focal length scales with 1/$\sqrt{P_{\rm AMB}}$, where $P_{\rm AMB}$ is the ambient pressure. The advantages of this model were that it is simple, physical and analytic. The disadvantages are that the clearing out of this hollow cone is hypothetical and that despite the scouring of the cone there is no deceleration of the jet. He evidently supposed the jets to have been launched recently, into the bubble of an SNR of about 50 pc radius. For the precession cone to have been collimated at this radius requires $P_{\rm AMB}\sim{10^{-11}}$  erg cm$^{-3}$. \citet{Eichler1983} suggested that this is reasonable within an SNR of radius $\sim$ 50 pc and so it is~\citep{CuiCox1992,Thorntonetal1998}.

Subsequent computations by \citet{PeterEichler1993} using the hydrodynamic code VH-1 showed that a diverging hollow conical jet can be collimated for a divergence angle rather greater than 5$^{\circ}$, thus verifying Eichler's conjecture and indicating conditions required for its validity. Their computations could never reproduce the expanding helices of Fig.~\ref{fblundell} because they approximated a diverging and precessing jet by a diverging hollow cone. However, these calculations suggest that jets both diverging and precessing might clean out the interior of the precession cone, collimation arising from the resulting pressure differential.

\subsection{Precessing jets in 3 dimensions}

Precessing jets, treated as cylindrical, were followed as they propagated through the ambient medium by~\citet{Zavalaetal2008}, using the 3D version of the YGUAZ'U-A code. They modelled not only the jets but also the supernova explosion forming the environment in which the jets propagate. Their most ambitious computation in fact modelled a supernova remnant expanding in a Galactic medium in which the density grew as the Galactic plane is approached, responsible for the greater extent of the east lobe in comparison to the west, which dips into the plane of the Galaxy. Qualitatively, they obtained a rather good description of the coarse features of the morphology of W 50 for a precession angle of 10$^{\circ}$ (their Fig.~6). But the precession angle of the real SS 433 is (currently) 20$^{\circ}$. They remark that ``it is of key importance to propose and test a model that can account for the early re-collimation of the jets".

A similar but perhaps even more ambitious sequence of computations is discussed in~\citet{Goodalletal2011b}. They used the 3D FLASH code and modelled jets. Initially jets were propagated into uniform media; later into the interior of a supernova remnant, with a density gradient in the pre-supernova medium. In most instances the precessing jets were cylindrical. They did not follow a model with 10$^{\circ}$ very far and concluded that the only way to reproduce the morphology of W 50 is to invoke several outbursts from the central engine. The cylindrical precessing jets do not diverge; while the FLASH code reveals some degree of collimation, it is not enough. \citet{Goodalletal2011b} find that the morphology of W 50 can be reproduced by a supernova explosion in the local medium and subsequent interaction with the precessing jets of SS 433, but because their jets are not significantly collimated the correct morphology requires three separate outbursts, first conical, secondly cylindrical jets and finally the precessing jets observed today. (For their model with expanding fireballs, \citet{Goodalletal2011b} have a divergence of about 0.01 rad only~\citep{Blundelletal2007}.)

\subsection{A model that achieves collimation}

\begin{figure}[h]
      \centering
      \includegraphics[width=8cm]{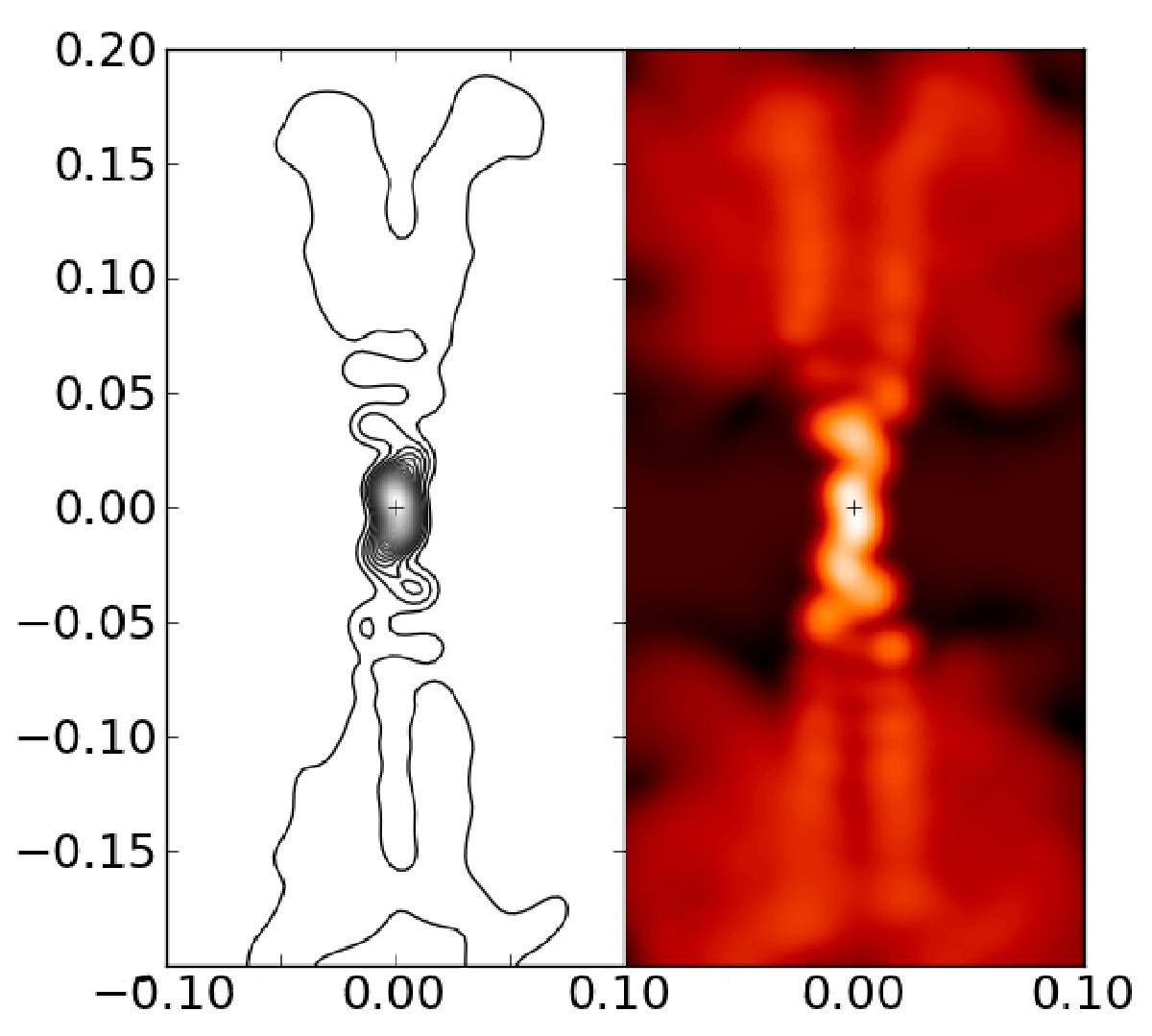}
      \caption{A computed radio map, under VLA conditions, for the jets of SS 433 evolving into the ambient medium of MB2 (their Fig.~9). The scales are in pc and the marked change in morphology at 0.07 pc is particularly clear in the right hand panel.}\label{fMB}
    \end{figure}

The most recent attempt to explain collimation follows numerically the evolution of a precessing jet that is itself diverging as it propagates into a uniform medium; that is, the precessing jet is itself a solid cone~\citep[hereafter MB1, MB2]{MBetal2014,MBetal2015}. The pictures of the jets propagating that are constructed from the computations with code MPI-AMRVAC are very striking (Figs.~1, 6, 9 in MB2). The jets decelerate, the pitch of the helices collapsing until, after about four turns (precessional periods), the morphology suddenly changes and the expanding helices turn into propagating hollow cylinders, see Fig.~\ref{fMB} of this paper, taken from Fig.~9 of MB2. In their computations, this transition occurs at a collimation distance of $\sim$0.07pc from the source of the jets. This apparently solves the collimation problem. There is a second feature emerging from those computations which is potentially of almost equal importance: the head of the propagating hollow cylinder travels outward along the precession axis at a speed $\sim$0.02c, rather than 0.24c as at launch. In Fig.~\ref{fMB} the head of the propagating cylinder has reached 0.18 pc after some 30 years but collimation took no more than $\sim$ 6 years. The slow speed suggests that the lack of any detected motion of filaments in the radio image of the ears of W 50 ($< 0.04$c, ~\citet{Goodalletal2011a}) can be understood in terms of deceleration from the early speed of 0.26c and also implies that the jet start up need not have occurred only a few thousand years ago, long after the original supernova explosion (the so-called latency problem, \citet{Goodalletal2011b}). These features are extremely promising, but the helices of the real SS 433 jets do not collimate within 0.1 pc - compare Fig.~\ref{fMB} with Fig.~\ref{fblundell}.

    While the description in the model of the jets of SS 433 seems realistic, the parameters describing the medium in which the jets propagate are very far from being an acceptable description of the true environment of SS 433. It is therefore of some importance to determine to what extent the promising features of the MB computations can survive a transition to more realistic parameters for the ambient medium. In the absence of further numerical computations, we have devised recipes for how the collimation distance and head velocity scale as the density and pressure of the ambient medium are changed: on scaling the solution of MB, their results for the relationship between W 50 and SS 433 survive.

    \begin{figure}[h]
      \centering
      \includegraphics[width=7cm]{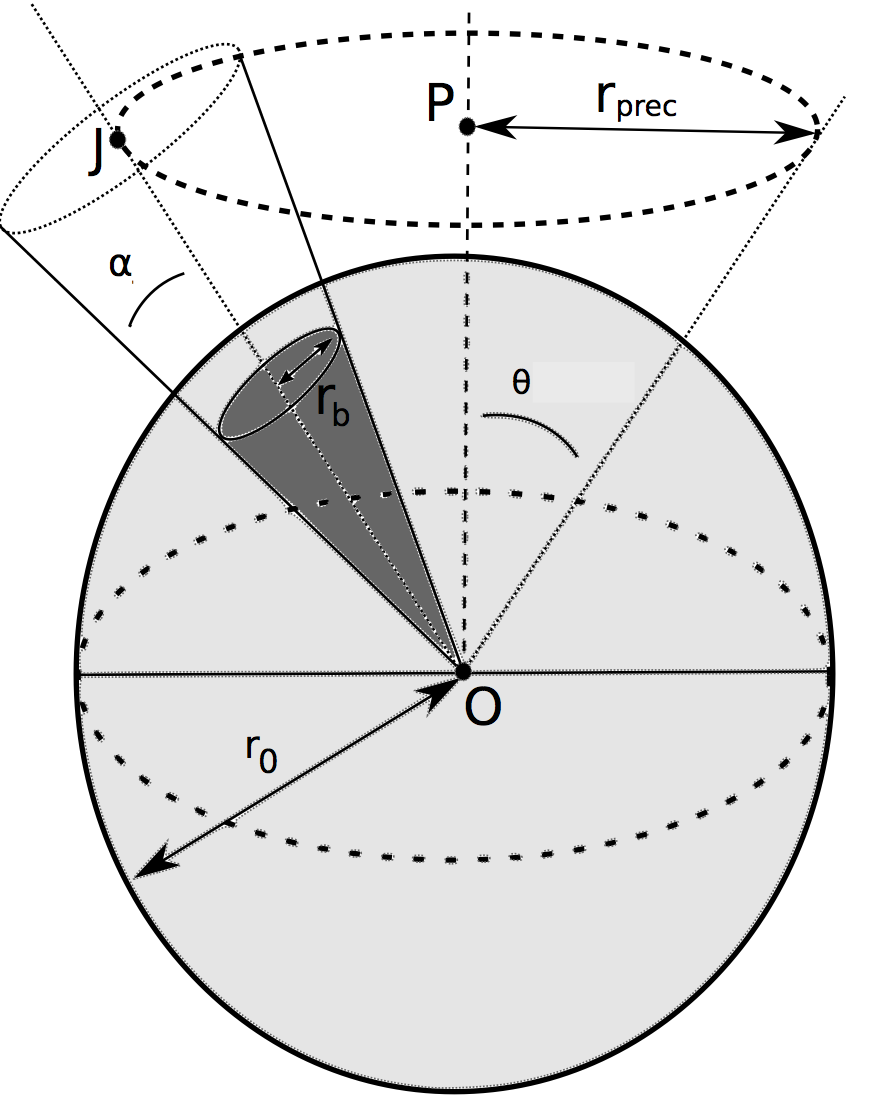}
      \caption{A schematic picture of the launch of a MB model SS 433 jet (adapted from Fig.~1 of MB1). Starting conditions for the diverging solid conical jet  that precesses about the axis OP are defined on the sphere of radius $r_{\rm 0} $ (see text).}\label{fgeometry}
    \end{figure}

The properties of the jets injected into the ambient medium in the computations of MB1, MB2 are easily understood with reference to Fig.~\ref{fgeometry}. Each jet is a solid cone diverging with a cone angle $\alpha$ of 5$^{\circ}$. The opening angle accounts not only for the intrinsic spread of the material on injection but also for the fast jitter caused by nutation. It is this solid conical jet that precesses with precession angle $\theta$ of 20$^{\circ}$. The starting conditions are defined on a sphere of radius $r_{\rm 0}$ (0.008 pc). At this distance the density $\rho_{\rm 0}$ is $2.58\times 10^{-22}$ gm cm$^{-3}$ and the radial velocities $v_{\rm 0}$ 0.26c on injection into the medium. The jet itself has a cross-sectional radius at $r_{\rm 0}$ denoted by $r_{\rm b}\approx r_{\rm 0}\tan \alpha$. Correspondingly, each jet has a kinetic luminosity $L$ (given by $0.5\rho_{\rm 0}v_{\rm 0}^{3}{\pi}r_{b}^{2}$) of $\ 10^{39}$ erg s$^{-1}$ and each jet transports 0.5 $10^{-6} M_\odot$ yr$^{-1}$.   

There is nothing unreasonable about these numbers and they are adopted for present purposes. However, the ambient medium into which the jets  were launched was taken as having a density of 5 $m_H$ cm$^{-3}$ and pressure $7.5\times 10^{-6}$ erg cm$^{-3}$. If the pressure is thermal, the corresponding temperature is $\sim 10^{10}$ K. The inapplicability of these numbers, which seem unrealistic, is in fact established directly from observation. The evolution with time of the helix traced out by the MB jets shows the pitch of the helix decreasing over several turns and almost vanishing in the fourth turn (Fig.~\ref{fMB}). VLA observations~\citep{BlundellBowler2004,Belletal2011} cover several turns along the helix and show no sign at all of a decreasing helix pitch (Fig.~\ref{fblundell}). The real values of ambient density and pressure within W 50 must be far lower than assumed in MB1, MB2. We have found that the principal features of the MB computations can be scaled rather simply to other values of pressure and temperature.

\section{Scaling relations}
\subsection{The need for and nature of scaling relations}
The collimation distance in the true environment of SS 433 will be far greater than that calculated in MB1, MB2. It is less obvious what the effect of more realistic parameters on the advance speed of the head of the cylindrical jet will be. How do we scale from the results obtained from the computations of MB2 to obtain values for different density and pressure? The way to proceed is suggested by equation (2) of MB2. It was conjectured that the morphology transition happens ``where the ram pressure of the jet elements is matched by the pressure that this element will encounter from the medium". Neglecting the Lorentz factor for the modest relativistic velocities involved, this condition yields Eq.~(2) of MB2 in the form
\begin{equation}
\frac{{\rho_{\rm 0}}{v_{\rm 0}^2}{d_{\rm 0}^2}}{d_{\rm 1}^2} =  P_{\rm ISM}                               \,, \label{eq1}
\end{equation}
where ${\rho_{\rm 0}}$, the density at injection into the medium, has been diluted by the square of the ratio of the distance from the origin at injection ($d_{\rm 0}$, equal in value to $r_{\rm 0}$) and the distance to collimation $d_{\rm 1}$. The quantity $P_{\rm ISM}$ in~(\ref{eq1}) refers to the pressure of an `interstellar' medium: to avoid possible confusion we refer rather from hereon to an ambient medium with pressure $P_{\rm AMB}$.The physics suggested seems not implausible and apart from an unknown dimensionless constant this would follow from application of the method of dimensions. In MB2 it was claimed that for the parameters in that paper the expression~(\ref{eq1}) yields for the collimation distance $d_{\rm 1}$ a value of 0.068 pc, in impressive agreement with the computed value of where the morphology changes.

This claim was the result of an unfortunate mistake, for in evaluating the expression, the assumed density of the ISM was inserted for ${\rho_{\rm 0}}$. When the correct jet value is used, the collimation distance is a factor of about 5 larger~\citep[erratum to MB2]{MBCor2017}. Nonetheless, it might well be the case that the collimation distance $d_{\rm 1}$ scales as 1/$\sqrt{P_{\rm AMB}}$. Collimation must occur before the jet penetrates the expanding shell; this then requires $P_{\rm AMB}> 10^{-11} $ erg cm$^{-3}$~\citep{Eichler1983}.

Similar arguments yield scaling for the velocity of the jet head, once the hollow cylinder has been formed. The speed of the jet head calculated as in the one-dimensional approximation of~\citet{Martietal1997} seems relevant. For an expanding jet this may be dubious, but on sounder footing for the cylindrical jet formed after the change in morphology. The argument is that the momentum along the precession axis, arriving over the area perpendicular to that axis, should drive forward the material in the ambient medium at the speed of the head. This will be discussed in more detail below, but the result that emerges is that the head speed scales with $\sqrt{T_{\rm AMB}}$, the square root of the temperature of the ambient medium. For the (eastern) jet of SS 433 to penetrate the shell and protrude through it to a distance equal to at least a shell radius requires $T_{\rm AMB }\sim {10^{8}}$ K.  A value $\sim{10^{10}}$ K may be unrealistic but $\sim{10^{8}}$ K is not, for such temperatures have been calculated for within the shocked gas filling the expanding bubble of a young supernova remnant~\citep{CuiCox1992,TrueloveMcKee1999}, see section~\ref{ssnr} of the present note. Their importance now being clear, details of the proposed scaling relationships and their physical origins are discussed in sections~\ref{scol} and \ref{svel} below.

\subsection{Scaling for the collimation distance}\label{scol}

There is reason to look further into the physical relevance of Eq.~(\ref{eq1}) rather than relying on a purely dimensional argument. In the simple dynamical model of~\citet{Eichler1983} collimation comes about through the external pressure squeezing his conical representation of the precessing jets, the focal length scaling with $\sqrt{L/{v_{\rm 0}}{P_{\rm AMB}}}$, where $L$ is the kinetic luminosity (power) of the jet. The collimation length from Eq.~(\ref{eq1}) has the same scaling, and it is also used by~\citet{Eichler1983} to normalize the distance scale in his analytical argument, so this scaling is simply applying the method of dimensions, quite independent of any dynamical model. We therefore adopt Eq.~(\ref{eq2a}) below, where Eq.~(\ref{eq1}) is modified by an unknown quantity $F$ that might be a function of the angle $\theta$ of the precession cone. For example, the conjecture that the pressure should be equated to that component of momentum flux perpendicular to the precession axis would give $F(\theta)\sim{\sin^2\theta}$. Equation~(\ref{eq2a}) below is directly related to Eq.~(\ref{eq1}); and it can be written alternatively as Eq.~(\ref{eq2b}) expressing the same physics without reference to the numerical nozzle in the computer model.
\begin{subequations}
\begin{align}
\frac{{\rho_{\rm 0}}{v_{\rm 0}^2}{F(\theta)}{d_{\rm 0}^2}}{d_{\rm 1}^2} =  P_{\rm AMB} \,, \label{eq2a}\\
\frac{2 L F(\theta)}{v_{\rm 0}{\pi}{d_{\rm 1}^2}{\tan^2\alpha}}= P_{\rm AMB} \,.\label{eq2b}
\end{align}
\end{subequations}
The function $F(\theta)$ is not known a priory, but it can be expected to grow as the ratio of $\theta$ to $\alpha$ grows. The reason is that the larger this ratio the harder it is to scour out the precession cone of the jets~\citep{Eichler1983,PeterEichler1993}. For the present case, the computations in MB2 give $d_{\rm 1}\sim$ 0.07 pc and hence from~(\ref{eq2a}) a value for $F(20^{\circ})\sim 0.05$, but we do not know in detail how it varies with $\theta$, the angle between the precessing jets and the precession axis. For this paper it does not matter, because we only scale with the temperature and pressure of the ambient medium, keeping the precession cone angle to the value for SS 433.

The model of~\citet{Eichler1983} suggested how collimation might come about and presented a useful scaling relation for focal or collimation lengths but did not contain any deceleration. The subsequent computations by~\citet{PeterEichler1993} demonstrated collimation for diverging conical jets. The marked and abrupt morphological changes that follow the expanding helices of precessing conical jets (as in Figs.~1, 6, 9 of MB2) had not been previously suspected.

\subsection{Scaling for the head velocity}\label{svel}

The velocity with which the head of the hollow cylindrical jet propagates after collimation is of great importance in understanding the ears of W 50. It is easy enough to construct an expression that is dimensionally correct and has some physical justification. For a cylindrical jet propagating through a uniform medium, the one-dimensional model of~\citet{Martietal1997} has been adapted in the following way. The cylindrical jet is assumed to commence in the plane normal to the precession axis at the collimation distance $d_{\rm 1}$ and the momentum there delivered must balance the momentum transferred to the medium by the advancing head. The rate at which momentum in the original (solid conical) jet passes any plane normal to the jet axis is $\rho_{\rm 0}{v_{\rm 0}^2}\pi(d_{\rm 0}\tan\alpha)^2$  or with the parameters employed in Eq.~(\ref{eq2b}) simply $2L/v_{\rm 0}$. Because of the precession, at the collimation distance $d_{\rm 1}$ the component along the precession axis is spread over a base area normal to that axis, see Fig.~\ref{fMB}. With a jet both diverging and precessing, this base area is proportional to $d_{\rm 1}^2$. If this area is a disk, then it is $\sim\pi(d_{\rm 1}\tan\theta)^2$; if the base were better treated as an annulus (see Fig.~7 of MB2) the area would then be $\sim {2\pi}{d_{\rm 1}^2}{\tan\alpha}\, {\tan\theta}$. For the derivation of a scaling relation only the dependence on $d_{\rm 1}^2$ is of importance.

Applying the prescription of~\citet{Martietal1997}, assuming a disk like base for the collimated jet, we obtain an equation for the speed $V$ at which the head propagates. In terms of the two sets of parameters we have
\begin{subequations}
\begin{align}
{\rho_{\rm 0}}{(v_{\rm 0}-V)^2}\frac{d_{\rm 0}^2}{d_{\rm 1}^2}\frac{\tan^2\alpha}{\tan^2\theta} =  {\rho_{\rm AMB}}{V^2}  \,, \label{eq3a}\\
{(1-\frac{V}{v_{\rm 0}})^2}\frac{2L}{\pi v_{\rm 0}{d_{\rm 1}^2}{\tan^2\theta}} = {\rho_{\rm AMB}}{V^2} \,. \label{eq3b}
\end{align}                                      
\end{subequations}
Here $\alpha$ is the (small) cone angle of the jet itself rather than of the precession cone; the latter is $\theta$; and $\cos\theta \sim 1$. The quantity $\rho_{AMB}$ is the density of the material being accelerated by the beam. 

Equations~(\ref{eq2a},\ref{eq2b}) can now be substituted into Eqns.~(\ref{eq3a},\ref{eq3b}), to yield
\begin{equation}
{(1-\frac{V}{v_{\rm 0}})^2} \frac{P_{\rm AMB} {\tan^2\alpha}}{{\tan^2\theta} {F(\theta)}} =   {\rho_{\rm AMB}}{V^2} \,, \label{eq4}
\end{equation}
where $F(\theta)$ is the function introduced in Eq.~(\ref{eq2a}). This equation is a rather exact analogue of Eq.~1 in~\citet{Martietal1997}, because a little calculation shows that the quantity $P_{AMB}\tan^2\alpha/{v_{\rm 0}^2}\tan^2\theta{F(\theta)}$ is just the density of material in the collimated cylindrical jet carrying the momentum of the original precessing jet.

If the head velocity is small compared with the launch velocity of the jets $v_{\rm 0}$ then, given the angles, the head velocity $V$ is determined only by the ratio $P_{\rm {AMB}}$/$\rho_{\rm AMB}$ and this ratio, apart from numerical factors, is the environmental temperature at the collimation distance initiating the cylindrical morphology. Hence, $V$ is closely related to the speed of sound in the ambient medium. That is, the speed of advance of the head of the (cylindrical) collimated jet does not depend on $ \rho_{AMB}$ and $P_{AMB}$ separately, only on the temperature of the ambient medium. This is a new and previously unsuspected result. In turn, the observed head velocity $V$ determines what the temperature must be in the environment at initiation of the cylindrical phase. With the parameters in MB2, their hydrodynamic computations gave $V=0.02c$. In view of the uncertainty in the appropriate base area entering into Eq.~(\ref{eq3a}), it is remarkable that Eq.~(\ref{eq4}) yields $0.03c$ with $F\sim0.05$ as determined through Eq.~(\ref{eq2a}). These speeds for the advancing head in the MB computations are indeed close to the speed of sound in the assumed ambient medium and here the one dimensional model of~\citet{Martietal1997} becomes inadequate. The scaling relation we have obtained from Eq.~(\ref{eq4}) is valid nonetheless.

For low head speeds, equations~(\ref{eq3a},\ref{eq3b}) need modification by adding to the right hand side the ambient pressure $P_{\rm AMB}$. This modification propagates into Eq.~(\ref{eq4}) and in calculating $V$ it contributes another term dependent on $P_{\rm {AMB}}$/$\rho_{\rm AMB}$. Thus, $V$ again scales with the square root of the temperature of the ambient medium.   In MB1, MB2 this temperature is the unrealistic value of $\sim{10^{10}}$ K.

Images of W 50 (such as Fig.~\ref{fdubner}) show that if the jets inflated the ears, then material from the early days of the jets must have covered $\sim$ 100 pc. The nebula is probably no older than $\sim10^{5}$ yr. The corresponding speed for the jet head is $\sim0.003c$; scaling then yields an ambient temperature at collimation of $\sim  10^{8}$ K.

\section{Supernova remnant models}\label{ssnr}
\subsection{General aspects}

The age of W 50 is presumed to be the time elapsed since the supernova explosion that also left the compact object in SS 433. A knowledge of the age is of course crucial to estimating the speed with which the heads of the jets trundle through the shocked medium within the expanding W 50. At the moment this age can only be obtained for W 50 from modelling the supernova explosion. To evaluate the applicability of the MB model, the pressure and temperature within the expanding remnant are also needed, as a function of both radius and time.
The evolution of a supernova remnant depends on the energy released in the explosion and on the characteristics of the medium in which the explosion occurs, initially as though that medium can be ignored (free expansion). When the blast wave has swept up a mass roughly equal to the mass ejected in the explosion, the evolution transitions from a radius growing linearly with time $t$ to its radius growing as $t^{2/5}$, the so-called Sedov-Taylor phase. Expansion of the SNR through both these phases is treated both analytically and numerically by~\citet{McKeeTruelove1995}. For reasonable parameters the age of a SNR of radius $\sim$50 pc is $\sim10^{5}$ years.  Modelling of supernova remnants yielding ages, pressures and temperatures within the cavity can be found in~\citet{CuiCox1992,Thorntonetal1998,KimOstriker2015}. These works report directly blast wave radii, temperatures and pressures and are broadly in agreement. There is no difficulty in finding pressures in a young SNR bubble that will collimate the precessing jets where the temperature is yet $\sim 10^{8}$ K. We take as an example~\citet{Thorntonetal1998}.

\subsection{A particular example}

Figs.~4 and 5 of~\citet{Thorntonetal1998} display a number of properties of their SNR as a function of radius, at relevant times of 9810 years and 1.27 $10^{5}$ years after an explosion releasing $10^{51}$ ergs into an ambient medium of density 0.133 $m_H$ cm$^{-3}$. The shell radii at these times are respectively 20 pc and 50 pc. If W 50 is correctly described by this model, then its age is little over $10^{5}$ years. At $\sim10^{4}$ years, the pressure within the shell is fairly uniform at $10^{-9.4}$ erg cm$^{-3}$ and at $\sim 10^{5}$ years $10^{-10.8}$ erg cm$^{-3}$. The corresponding distances to collimation are thus, scaling the 0.07 pc of MB2, 9 pc and 44 pc. In either case collimation would take place within the cavity.
The temperature of the material in the cavity is also shown as a function of radius. At 9810 years the temperature is $>10^{8}$ K out to a radius of 5 pc and is $>10^{7}$  K out to the shell, at a radius of 20 pc. At the much later time of $\sim10^{5}$years, the temperature falls below $10^{7}$ K at 25 pc. 

Thus if the supernova occurred $\sim10^{5}$ years ago and the jets initiated $\sim10^{4}$ years later, this model of a supernova remnant comes very close to the conditions necessary for the structure of W 50 to be well described by the model of MB1, MB2, together with the scaling relations proposed here. If the W 50/SS 433 system is younger, the head of the cylindrical jet would have had to propagate faster. If the jets were launched earlier than 9810 years after the explosion, environmental pressures and temperatures would be higher (see~\citet{CuiCox1992}). It all seems possible.

As the pressure and temperatures within the SNR cavity drop, the collimation distances grow slowly with time and the speeds of the cylindrical jets drop even further. It is tempting to speculate that this might lead to late launched material contributing hollow cylindrical jets of larger radii, easing the blending of the spherical shell of W 50 and the jet driven protrusions.

\subsection{A counter example?}
It would not be proper to end this section without drawing attention to a computation which disagrees with the results discussed above in $4.2$. Those results all attribute an age of $\sim 10^{5}$ years to an SNR of radius $\sim45$ pc. In Fig.~3 of~\citet{Goodalletal2011b} are illustrated various stages in the expansion of an SNR with parameters close to those in~\citet{Thorntonetal1998}. The age of their SNR at a radius of 45 pc is $\sim$18 000 years, about one fifth of other results. If this were true, the temperature at initial launch of the jets would need to be several times $10^{9}$ K and the reasoning presented above in section $4.2$ would fall apart. \citet{Goodalletal2011b} imply that the reason for this discrepancy is that other calculations ignored the free expansion phase covering $\sim$5 pc. Now 5 pc covered very fast, out of 45 pc, is not obviously going to reduce the total time by a factor of 5. The paper by~\citet{McKeeTruelove1995} covered free expansion followed by transition into Sedov-Taylor expansion. They present both computations and analytic approximations and for the parameters used in~\citet{Goodalletal2011b} their formulae give the time taken to reach a radius of 45 pc to be $\sim$ 60 000 years. It would appear that explicit inclusion of free expansion is not responsible for the factor 5 discrepancy between the age according to~\citet{Goodalletal2011b} and the ages from many other calculations. For a radius of 45 pc at 18 000 years the formulae of~\citet{McKeeTruelove1995} require a pre-supernova density of 0.005 $m_H$ cm$^{-3}$.

\section{Discussion}

The purely analytic treatment of~\citet{Eichler1983} found collimation of precessing jets through the pressure of the ambient medium. That model contained no deceleration, despite appealing to sweep-out of the conical cavity by the precessing jet. \citet{PeterEichler1993} treated a diverging hollow conical jet and showed that for sufficient interior sweeping such jets can indeed be collimated; they express strongly the opinion that sufficient divergence within the precessing jets is crucial. This may well be so; in the ambitious paper of~\citet{Goodalletal2011b} the precessing jets are cylindrical and do not diverge; while their FLASH code reveals some degree of collimation, it is evidently not enough (see also~\citet{Zavalaetal2008}).
 
We note that supersonic, solid conical jets can show a self-similar structure with a cavity bounded by swept-up ambient gas on the larger scales~\citep{Falle1991}. Within these models, jet reconfinement shocks can be observed depending on the precise density distribution with distance from the source region. This is rather different from the smooth dynamical changeover from a precessing jet to a cylindrical hollow jet as observed after some windings in the MB computations. It is clear that the uniform surrounding medium assumption adopted in MB must be altered to one where the precessing jet enters a true SNR environment, as was attempted in the~\citet{Goodalletal2011b} set of simulations. This can be subject of future computational work, where the main challenge will be to bridge the factor 1000 in scales as mentioned between Fig.~\ref{fblundell} and Fig.~\ref{fdubner}.
Only such computational approaches can handle the additional complexities that abound when jets interact with spatio-temporally dependent environments. Indeed, our simple ambient medium parametrization is understood to vary with space and time, as a supersonic jet will sweep up and shock matter that it encounters. Moreover, when the (collimated) jet ultimately protrudes through the SNR remnant shell, it will change to see the ISM conditions as ambient.

The effective divergence employed by MB1, MB2 was set to about 0.1 rad. For the case studied by~\citet{PeterEichler1993} a divergence of this order was found necessary for collimation to occur. This $5^{\circ}$ divergence in MB1, MB2 was introduced largely to account for the nodding of the jets due to nutation. The nodding is real (vividly illustrated in Fig.~3 of~\citet{Blundelletal2007}); the jets of SS 433 do effectively diverge and so, precessing, can metamorphose into hollow cylinders. 

The changeover from truly helical jets 
to hollow cylinders can be seen in MB2, their Fig.~7. 
An indication of fine-structure on the cylindrical part can be detected in cuts across the jet precession axis, beyond the point of collimation into this cylinder. This may be related to the recent discovery of stability loss of highly supersonic, solid conical jets as studied recently by~\citet{GourgouliatosKomissarov2018} in the context of active galactic nuclei jets. There, the solid conical, highly relativistic jets showed the expected pressure-mediated recollimation beyond the reconfinement point, accompanied by a novel turbulent changeover beyond this point due to centrifugal instability. This centrifugal effect comes about from the curved path taken by parcels if acted upon by external pressure, as already analysed in~\citet{Eichler1983}. In the MB1, MB2 computations,
individual parcels are launched radially in the diverging jet nozzle, but they then seem to align in a helix due to the imposed slow precession of the jet injection region illustrated in Fig.~\ref{fgeometry}. They collectively undergo deceleration due to the increased interaction surface with the ambient medium, and ultimately this helix blends into a hollow cylindrical structure. The computations show no true refocusing of this structure to the precession axis, but fine-structure does appear all along the cylindrical jet. This specific SS 433 morphology, and the overall parameter regime (where only mildly relativistic flow is involved) is very different from the configuration studied in~\citet{GourgouliatosKomissarov2018}, and future simulations are needed to address such questions as centrifugal instability.

\section{Conclusions}

The computations of MB1, MB2 demonstrate both collimation and deceleration of the jets. Subject to the scaling relations proposed here, the following conclusions can be drawn. First, precessing jets can be collimated by ambient pressure; the cone angle close to the central engine of SS 433 could always have been $\sim20^{\circ}$. Secondly, the cylindrical jets formed at collimation have heads propagating at a velocity closely related to the speed of sound in the ambient medium, $\sim0.003c$. This could explain lack of visible motion of filaments in the ears and long latency is not required. Thirdly, this speed corresponds to temperatures likely to be found within a young SNR bubble.  W 50 could have been generated by a supernova explosion, followed not long after by the jets that eventually inflated the ears.
It might be that the various problems concerning the relationship between the nebula W 50 and the microquasar SS 433 are now solved.

\begin{acknowledgements}
MGB thanks DRB for TeXpertise
\end{acknowledgements}

\bibliographystyle{aa}
\bibliography{bowler}

\section*{Glossary}

The geometrical descriptions of jets can be ambiguous and confusing. In this glossary we attempt to present clearly the meanings of the terms employed in this paper.

In most instances, the geometric description applies really to the trajectories material in a jet would follow after launch into vacuum. Thus a solid cylindrical jet is the name given to material launched uniformly perpendicular to the surface of a disk. In the absence of any interference (internal or external pressures) it would maintain this cross section indefinitely, the material moving parallel to the surface of the cylinder.

   A hollow cylindrical jet is material launched uniformly from the surface of an annulus and perpendicular to it; the surfaces are those of two co-axial cylinders of different radii. Macaroni as opposed to spaghetti. The description of originally precessing jets after collimation as hollow cylinders (Fig.~\ref{fMB}) does not quite match the technical definition just given but should be clear.

  The categories of conical jets are even more confusing. Consider first the figure formed by two geometrical cones of equal cone angle but displaced one from the other along the common axis. For a hollow conical jet the material is launched between the two cones and parallel to the conical surfaces. If two geometric cones have slightly different cone angles and the axes and vertices are common, the material launched between them constitutes a diverging hollow conical jet. A solid conical jet is not hollow; rather material is launched uniformly over the solid angle subtended at the vertex.

   All the idealised geometrical jets have rotational symmetry about an axis and if a jet is precessing it is this axis of symmetry that precesses. If a precessing jet is approximated by a conical jet, the axis of symmetry of the conical jet is of course the precession axis for the original.

\end{document}